\documentclass[twocolumn,epjc3]{svjour3}

\RequirePackage[T1]{fontenc}

\smartqed 
\RequirePackage{graphicx}
\RequirePackage{amssymb}
\RequirePackage{mathptmx}     
\RequirePackage{flushend}
\journalname{Eur. Phys. J. C}

\begin{document}

\title{Neutron oscillations to parallel world: earlier end to the cosmic
ray spectrum?} \author{Zurab Berezhiani\thanksref{e1,addr1,addr2} 
\and
Askhat Gazizov\thanksref{e2,addr3}
}

\thankstext{e1}{e-mail: zurab.berezhiani@aquila.infn.it}
\thankstext{e2}{e-mail: askhat.gazizov@desy.de}

\institute{Dipartimento di Fisica, Universit\`a dell'Aquila, Via Vetoio,
67100 Coppito, L'Aquila, Italy \label{addr1} \and
INFN, Laboratori Nazionali Gran Sasso, 67100 Assergi, L'Aquila, Italy
\label{addr2}
\and
DESY Zeuthen, Platanenallee 6, D-15738 Zeuthen, Germany \label{addr3}
}

\date{Received: date / Accepted: date}

\maketitle

\begin{abstract}% 
Present experimental data do not exclude fast oscillation of the neutron
$n$ to its degenerate twin from a hypothetical parallel sector, the so
called mirror neutron $n'$. We show that this effect brings about
remarkable modifications of the ultrahigh-energy cosmic ray spectrum
testable by the present Pierre Auger Observatory (PAO) and Telescope
Array (TA) detector, and the future JEM-EUSO experiment. In particular,
the baryon non-conservation during UHECR propagation at large
cosmological distances shifts the beginning of the GZK cutoff to lower
energies, while in presence of mirror sources it may enhance the
spectrum at $E > 100$ EeV. As a consequence, one can expect a
significant reduction of the diffuse cosmogenic neutrino flux.
\end{abstract}%

There may exist a hidden parallel sector that is an exact copy of our
particle sector. One can imagine a theory based on a direct product 
$G \times G'$ of identical gauge factors with identical particle contents
which can emerge, e.g. in the context of $E_8 \times E'_8$ string
theory. As a minimal possibility, one can consider a case of two
Standard Model copies with $G=SU(3)\times SU(2)\times U(1)$ standing for
ordinary sector and $G'=SU(3)'\times SU(2)'\times U(1)'$ standing for
parallel sector. Alternatively, one can envisage some grand unified
extensions as $SU(5) \times SU(5)'$, etc. The Lagrangians of two worlds
can be rendered identical to each other, with all coupling constants
being exactly the same in both sectors, by introducing a discrete
symmetry $G\leftrightarrow G'$ under the exchange of two gauge systems
and of the respective matter fields.
 
A well-known example, coined as mirror world
\cite{Lee:1956qn,Kob:1966,Nishijima:1965zza,Blinnikov:1983gh,Foot:1991bp,Hodges:1993yb}, 
was introduced long time ago for interpreting parity as a 
discrete symmetry when our `left-handed' particles are exchanged with their 
{\it mirror} twins that are `right-handed'. Concerns about parity are
irrelevant for our following discussions which can be extended to a
parallel sector (or sectors) of any chirality. For us is important only
that each ordinary particle: electron $e$, proton $p$, neutron $n$ etc.\
may have a mass degenerate twin: $e'$, $p'$, $n'$ etc. These twin
particles must be sterile to our strong and electroweak interactions but
have their own strong and electroweak interactions among
themselves.\footnote{ In the following, for terminological simplicity,
we shall continue to call the particles of the `primed' parallel sector
as mirror particles, independently of their chirality. }

Mirror baryons can be viable as \emph{asymmetric} dark matter provided
that parallel sector has smaller temperature than the ordinary one, $T'
\ll T$
\cite{Berezhiani:2000gw,Ignatiev:2003js,Berezhiani:2003wj,Berezhiani:2005vv}.
On the other hand, once this condition is fulfilled, ${\rm B}\!-\!{\rm
L}$ and CP violating interactions among ordinary and mirror particles
can generate baryon asymmetries in both sectors
\cite{Bento:2001rc,Bento:2002sj}, naturally giving the relation
$\Omega'_{\rm B}/\Omega_{\rm B} \simeq 5$ between cosmological fractions
of the dark and visible matter
\cite{Berezhiani:2003xm,Berezhiani:2005ek,Berezhiani:2008zza}. Such
interactions can be mediated by heavy messengers coupled to both
sectors, as right handed neutrinos or extra gauge
bosons/gauginos~\cite{PLB98}. In the context of extra dimensions,
ordinary and mirror sectors can be modeled as two parallel 3-dimensional
branes and particle processes between them could be mediated by the bulk
modes or ``baby branes"~\cite{Gia}.

The same B or L violating interactions that lead to primordial
baryogenesis can also induce mixing phenomena between the ordinary
particles and their mirror partners. E.g.\ effective operator
$(1/M)l\phi l'\phi'$ ($\Delta {\rm L}=1$) between the ordinary/mirror
lepton and Higgs fields in the early universe gives rise to an efficient
lepto-baryogenesis mechanism for both sectors
\cite{Bento:2001rc,Bento:2002sj,Berezhiani:2003xm}, while at low
energies it induces the mixing between ordinary (active) neutrinos
$\nu_{e,\mu,\tau}$ and their mirror (sterile) partners
$\nu'_{e,\mu,\tau}$ \cite{Foot:1995pa,Berezhiani:1995yi} (see also
\cite{Foot:1991py,Akhmedov:1992hh}).\footnote{Mirror symmetry can be
spontaneously broken e.g. due to the difference of weak interaction
scales or grand unification scales between two sectors. Then the mirror
sector can be deformed to a shadow world with certain predictable
properties. Some phenomenological and cosmological implications of such
models were discussed in refs.\ \cite{BCV5,BCV6,BCV6A,BCV8}. }

Effective six-fermion interactions $(1/M)^5 (udd)(u'd'd')$, etc.\ 
($\Delta {\rm B}=1$) with the scale $M\sim 1-10$ TeV, involving the
ordinary ($u,d$) and mirror ($u',d'$) quarks of different families can
provide an efficient mechanism for primordial baryogenesis and dark
matter genesis, and in addition, can be testable at the LHC
\cite{Berezhiani:2005hv}. On the other hand, at low energies these
operators
induce the mass mixing $\varepsilon (\overline{n} n' + \overline{n}' n)$ 
between the neutron $n$ and its mass degenerate mirror twin $n'$, 
with $\varepsilon \sim \Lambda^6_{\rm QCD}/M^5$. Hence, in the vacuum
conditions $n$ and $n'$ must have a maximal mixing while the oscillation
time in the rest frame, $\tau_{nn'} = \varepsilon^{-1} \sim (M/10\, {\rm
TeV})^5$~s, can be much smaller than the neutron $\beta$-decay time
$\tau_{\rm d} \simeq 880$~s, and in fact it can be of the order of a
second \cite{Berezhiani:2005hv}. This seems rather surprising as it
regards the baryon number violating process, but it is not excluded by
the present experimental data.\footnote{ Other $\Delta{\rm B} =1$
effects as e.g.\ oscillation $\Lambda$--$\Lambda'$ between the hyperons
can be even faster. However, they are very short-living and these
effects should be more difficult to observe.} The key moment is that for
free neutrons $n$--$n'$ transition is affected by the Earth magnetic
field whereas for the neutrons bound in nuclei it is ineffective and so
the nuclear destabilization limits are irrelevant. In addition,
$n$--$n'$ oscillation with $\tau_{nn'} \sim 1$~s is not in conflict with
the astrophysical bounds from primordial nucleosynthesis, from the
neutron star stability, etc.\ 
\cite{Berezhiani:2005hv,Berezhiani:2006je,Berezhiani:2008bc}.\footnote{For
other theoretical works on $n$--$n'$ oscillation see \cite{Pokot,Nasri,Redi}.}

In the last years several experiments searched for $n-n'$ oscillation
via the magnetic field dependence of the neutron losses
\cite{Ban:2007tp,Altarev:2009tg,Serebrov:2007gw,Serebrov:2009zz}. Note,
however, that the lower bounds on $\tau_{nn'}$ reported by these
experiments (the strongest limit reads $\tau_{nn'} > 414 $~s
\cite{Serebrov:2007gw}) and adopted by the Particle Data Group
\cite{PDG:2011} become invalid if the Earth possesses a reasonably large
mirror magnetic field: the latter cannot be screened in the experiments
and it can drastically influence the probability of $n-n'$ oscillation
\cite{Berezhiani:2008bc}. On the other hand, if dark matter consists of
mirror particles, it is also plausible that the solar system and the
Earth itself may capture a significant amount of mirror matter due to
some feeble interactions between ordinary and mirror particles. The
Earth's rotation gives rise to circular currents in the captured mirror
matter which may induce a mirror magnetic field up to several Gauss
\cite{Berezhiani:2008bc}.

In the presence of non-zero mirror field $B'$ the probability of $n-n'$
oscillation has a non-trivial dependence on the ordinary magnetic field
$B$ and on its orientation relative to $B'$ \cite{Berezhiani:2008bc}. In
particular, if $B' > 0.05$~G, the existing experimental bounds do not
exclude $\tau_{nn'} < 10$ s or so. Moreover, the data acquired in the
experiment \cite{Serebrov:2009zz} provide a positive signal for
$n$--$n'$ oscillation with $\tau_{nn'}$ of few seconds. According to
critical analysis performed in ref.\ \cite{Berezhiani:2012rq}, these
data indicate that the ultra-cold neutron losses measured in magnetic
field depend on the magnetic field strength and its orientation. Namely,
the measurements performed at $B=0.2$~G show dependence on the 
magnetic field direction at more than 5$\sigma$ level whereas no effect is 
seen in the measurements performed at $B=0.4$~G. This anomaly can 
be interpreted in terms of $n-n'$ oscillation with $\tau_{nn'}=2-10$~s
provided that the Earth possesses a mirror magnetic field $B'\simeq
0.1$~G \cite{Berezhiani:2012rq}. This result, if confirmed by
forthcoming experiments, will have deepest consequences for fundamental
particle physics, astrophysics and cosmology.

In this letter we show that fast $n$--$n'$ oscillation can have
intriguing implications for the propagation of ultra-high-energy cosmic
rays (UHECR). In particular, it can cause significant modifications of
the spectrum at $E \gtrsim 10$ EeV which can be proved in future
experiments with high accuracy.

It is known that the cosmic microwave background (CMB) causes an abrupt
end in the cosmic proton spectrum, the so called Greisen-Zatsepin-Kuzmin
(GZK) cutoff \cite{Greisen:1966jv,Zatsepin:1966jv}. The cutoff energy
corresponds to the pion photoproduction threshold, $E_{\rm GZK}= m_\pi
m/2\varepsilon_\gamma \simeq 60$~EeV, where $m$ and $m_\pi$ are
respectively the nucleon and pion masses and $\varepsilon_\gamma \simeq
3 T \simeq 10^{-3}$ eV is an effective energy of relic photons,
$T=2.725$~K being the CMB temperature. The mean free path (m.f.p.) of
the proton, $l_{\rm s} \sim \langle \sigma(p\gamma \rightarrow N\pi)
n_{\gamma}\rangle ^{-1} \propto T^{-3}$, strongly depends on the energy.
One has $l_s \sim 5$~Mpc for $E > 300$~EeV but it sharply increases at
lower energies, e.g. $l_s > 100$~Mpc at $E < 60$~EeV. In each $p\gamma
\rightarrow N\pi$ scattering with one pion production the super-GZK
protons lose about 15--20\% of their energy, but at large energies, $E
\gg E_{\rm GZK}$, the energy loss by multi-pion production can
effectively reach 50\%.

The $p\gamma$-scattering has two main pion production channels, $p
\gamma \rightarrow p \pi^0$ and $p\gamma \rightarrow n \pi^+$, with
roughly comparable cross-sections. Conversion of the cosmic ray proton
into the neutron does not influence the propagation length, since
$n\gamma \rightarrow N\pi$ scatterings, $n\gamma \rightarrow n\pi^0$
($p\pi^-$), have nearly the same cross sections as $p\gamma \rightarrow
N\pi$ ones. In addition the $\beta$-decay $n \rightarrow p e
\tilde{\nu}_e$ converts the neutron back to the proton with practically
the same energy. Up to $E\simeq 0.5$~ZeV the decay length $l_{\rm
d}=\Gamma c\tau_{\rm d}$ ($\Gamma=E/m$ is Lorentz factor) is smaller
than $n\gamma \rightarrow N \pi$ scattering length. Hence, cosmic ray
carriers with $E \gg E_{\rm GZK}$ travel long distances transforming
from protons to neutrons and back suffering significant energy losses
which downgrade their energy to sub-GZK range. Yet, the baryon number in
the cosmic ray propagation is conserved.

In presence of $n-n'$ oscillation the situation changes drastically: the
produced neutron can now oscillate into a mirror one. If $\tau_{nn'}\ll
\tau_{\rm d}$ the oscillation length $l_{\rm nn'} = c\Gamma \tau_{nn'}$
is much smaller than $l_{\rm d}$ and $l_{\rm s}$ so that at these scales
oscillations may be averaged. The $n-n'$ transition probability was
calculated in \cite{Berezhiani:2005hv,Berezhiani:2006je} and in general
case in \cite{Berezhiani:2008bc}. For the cosmic neutron oscillations it
reads 
\begin{equation} % 
\label{PE} %
P(E) = \frac{1}{2[1 + ( \Gamma \omega \tau_{nn'} )^2]} = 
\frac{1}{2 + q ( E/100\,{\rm EeV})^2}, 
\end{equation} % 
where $\omega = \frac12 |\mu_n \Delta \mathfrak{B}|$ and 
$\Delta \mathfrak{B} = \mathfrak{B} - \mathfrak{B}'$, $\mu_n$ being 
the neutron magnetic magnetic moment and
$\mathfrak{B}$ and $\mathfrak{B}'$ being respectively the ordinary and
mirror magnetic fields at the cosmological scales, or more precisely
their transverse components. Factor $q = 0.45 \times (\tau_{nn'}/1~{\rm
s})^2 \times (\Delta \mathfrak{B}/1~{\rm fG})^2$, shows the efficiency
of $n-n'$ oscillation at $E \simeq E_{\rm GZK}$. Finally, $\beta$-decay
of mirror neutron $n' \rightarrow p' e' \tilde{\nu}'_e$ converts a
cosmic ray, being initially a proton, to a mirror proton.

The latter can be converted to ordinary proton via inverse chain of
reactions: $p'\gamma'\rightarrow n'\pi'$ scattering, $n'-n$ transition
and $n\rightarrow p e\tilde{\nu}_e$ decay. However, the propagation
length of mirror protons is much larger than that of ordinary ones,
$l'_s \gg l_s$, as far as the temperature of mirror CMB is smaller than
that of ordinary CMB, $T'/T = x \ll 1$. Namely, the Big Bang
nucleosynthesis imposes a robust upper bound $x < 0.5$ or so
\cite{Berezhiani:2000gw,Ignatiev:2003js,Berezhiani:2003wj}, but the
limits strengthen if one assumes that dark matter consists entirely of
mirror baryons. In this case the large scale structure and CMB power
spectrum require $x< 0.3$, while yet stronger limits as $x<0.2$ (or
$x<0.1$) arise by demanding that the Silk damping of mirror baryon
perturbations does not prevent the formation of normal (or dwarf)
galaxies 
\cite{Berezhiani:2000gw,Ignatiev:2003js,Berezhiani:2003wj,Berezhiani:2003xm}.

For the relic mirror photon number density and their average energy we
have $n'_\gamma = x^3 n_\gamma$ and $\varepsilon'_\gamma =
x\varepsilon_\gamma$. Thus m.f.p.\ of mirror cosmic rays is drastically
amplified, $l'_{\rm s}/l_{\rm s} \simeq x^{-3} \gg 1$, while the
threshold energy of $p'\gamma'\to N'\pi'$ increases as well, $E'_{\rm
GZK} \simeq x^{-1} E_{\rm GZK}$. So, the energy range $E_{\rm GZK}
\lesssim E \lesssim E'_{\rm GZK}$ acts for ordinary cosmic rays like a
sink where they disappear -- ordinary cosmic rays with $E > E_{\rm GZK}$
are converted to mirror ones, but the mirror ones may be converted (at
much lower rate) to ordinary ones only at $E > E'_{\rm GZK}$. The
dominant fraction of the cosmic rays produced in far distant sources
must escape to the parallel sector via $n-n'$ oscillation. However, if
there are powerful mirror sources the ordinary UHECR flux may be
increased at $E> E'_{\rm GZK}$ by the contribution from cosmic rays
originated in the mirror sector and converted to ordinary ones via
$n'$--$n$ oscillation.

In the presence of $n-n'$ oscillation, evolution of the four UHECR
number densities $U_i=U_i(E,t)$, $i=p,n,p',n'$, in the expanding
universe may be described by a system of coupled integro-differential
equations 
\begin{equation} 
\begin{array}{l} %
\frac{\partial U_i}{\partial t}= Q_i -3 H(t) U_i +\frac{\partial
\left[E(H(t)+\beta_i) U_i\right] } {\partial E} + \frac{m
D_{ij}}{E\tau_{\rm d}} U_j \\ %
- R_i(E,t) U_i + T_{ij}(E) \int \limits_E^{\infty} \! d\tilde{E} \,
W_{jk}(E, \tilde{E},t) U_k(\tilde{E},t), 
\end{array} %
\label{system}% 
\end{equation} 
where $H(t)$ is Hubble parameter. We assume that cosmic rays sources are
distributed homogeneously in space and their generation functions
$Q_i(E,t)$ may have cosmological evolution with time $t$
\cite{Berezinsky:2002nc}. Here $W_{jk}(E,\tilde{E},t)$ is the
probability density for a nucleon $k$ ($N=p,n$ or $N'=p',n'$) of energy
$\tilde{E}$ to transform via the pion-production scatterings off the
respective CMB ($\gamma$ or $\gamma'$) into a nucleon $j$ (again $N$ or
$N'$) with energy $E$. Hence, the mixed terms between two systems
vanish, $W_{NN'}=0$, whereas the relevant terms are calculated using the
cross sections of $p\gamma$ and $n\gamma$ processes which take into
account also the multi-pion production channels. The matrix $T_{ij}(E)$,
with $T_{pp}=T_{p'p'}=1$, $T_{nn}=T_{n'n'}=1-P(E)$,
$T_{nn'}=T_{n'n}=P(E)$ given by Eq.~(\ref{PE}) and with other elements
being zero, stands for transition probabilities due to $n-n'$
oscillation: the neutron $n$ produced in $N\gamma$-scattering, $N=p,n$,
promptly oscillates into $n'$ with a probability $P(E)$ and vice versa;
$R_i(E,t) = \int_0^E d\tilde{E} \sum_j W_{ji}(\tilde{E},E,t)$ stands for
probability of a nucleon $i$ with energy $E$ to disappear from the
energy range $dE$; factors $\beta_{p,p'}(E,t)$ take into account the $p$
and $p'$ energy losses due to $e^+ e^-$ pair production
($\beta_{n,n'}=0$). The matrix $D_{ij}$ takes into account $n$ and $n'$
$\beta$-decays transforming neutrons back into protons with practically
the same energy; here $D_{pn} = D_{p'n'} = 1$, $D_{nn} = D_{n'n'} = -1$
and all other elements are zero. In the absence of $n-n'$ oscillation
the system (\ref{system}) obviously splits into two independent sets of
equations for ordinary and mirror cosmic rays.

At all reasonable energies neutrons decay before their $n\gamma
\rightarrow N\pi$ scattering off CMB, so that $l_{\rm d} < l_{\rm s}\ll
l'_{\rm s}$ (needless to say that we take into account also the
multi-pion production). The relation $l_{\rm d} \ll l'_{\rm s}$ holds
very well for mirror neutrons, while for ordinary ones $l_{\rm d} <
l_{\rm s}$ is fulfilled only at $E < 500$~EeV. For this energy range the
initial proton, after $p\gamma \rightarrow n \pi^+$ scattering, with
following prompt $n-n'$ oscillation and neutron decay, instantly
transforms into a mirror proton, $p \rightarrow p'$ with probability
$P(E)$, and vice versa, $p' \rightarrow p$, neglecting the propagation
periods when a nucleon dwells in the mixed $n-n'$ state. This allows to
integrate out $n$ and $n'$ states and to reduce the system of 4
equations (\ref{system}) effectively to a system of two equations
describing evolution of just $p$ and $p'$. It should be noted that this
approximation is equivalent also to the approximation when one neglects
the difference between $p\gamma$ and $n\gamma$ cross-sections.

To facilitate the calculations, the latter system of two
integro-differential equations was reduced to an analog of a set of two
coupled Fokker-Plank type differential equations by expanding the
kernels of integrals in (\ref{system}) in series at $\tilde{E} = E$ up
to second derivatives. This method was proved to be valid by comparison
with Monte Carlo simulations in the case of propagation of ordinary
protons \cite{Berezinsky:2006mk}. Moreover, we assume $q \ll 1$ in
Eq.~(\ref{PE}), so that $P(E) = 1/2$. This holds for $\Delta
\mathfrak{B} \lesssim 1$~fG which is consistent with the limits on
extragalactic magnetic fields $\mathfrak{B} > 10^{-2}-1$~fG given in
ref. \cite{Taylor:2011bn}. There is no strong evidence for a presence of
larger magnetic fields in voids. The galactic magnetic field generation
mechanism from the density perturbations before the recombination
predicts seed magnetic fields at the scales larger than 1 Mpc smaller
than $10^{-4}$~fG \cite{Berezhiani:2003ik,Matarrese:2004kq}. However,
the data concerning the intergalactic magnetic fields are controversial
and there are some hints that they may be larger than 1 fG.
Nevertheless, at large scales ordinary and mirror magnetic fields can be
strongly correlated, so that their difference $\Delta \mathfrak{B}$ can
be small enough \cite{Berezhiani:2006je}. There is also a possibility of
the resonance MSW like $n$--$n'$ transitions if magnetic fields larger
than 1~fG have turbulent structure at scales less than 1
Mpc.\footnote{Let us remark, that rather large magnetic fields would be
necessary for the MSW-like transitions in the bigravity picture when
gravity is not fully universal between two sectors, e.g.\ when ordinary
and mirror sectors have separate gravities, forming one massless and one
massive graviton eigenstates \cite{Berezhiani:2009kv,Berezhiani:2009kx}.
}

\begin{figure} % figure 1
\begin{minipage}{\columnwidth}
\centering
\includegraphics[width=\columnwidth]{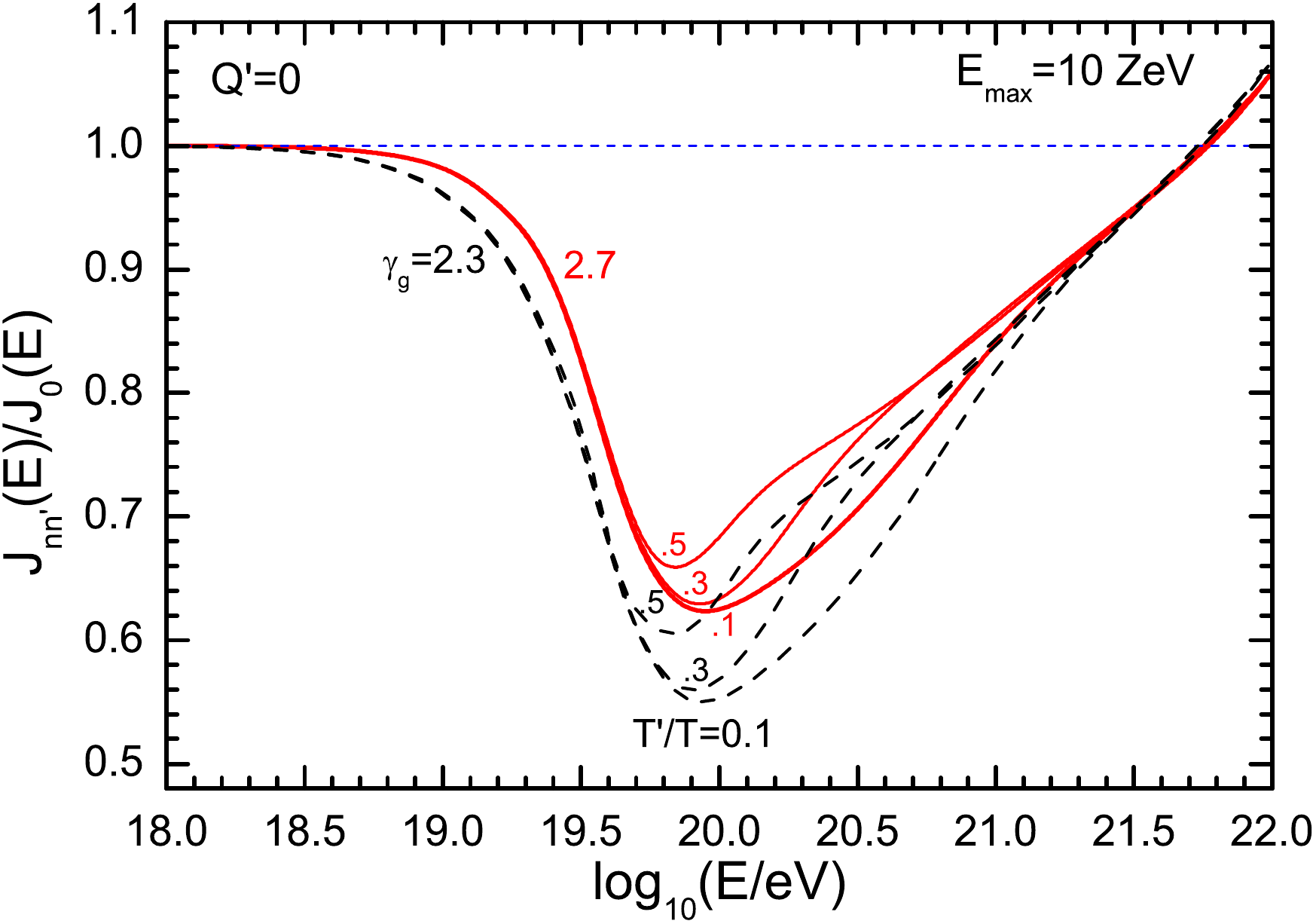}
\end{minipage}
\caption{Ratios of the UHECR spectrum modified by $n-n'$ oscillation,
$J_{nn'}(E)$, and standard GZK ({\it no oscillation}) spectrum,
$J_0(E)$, in the absence of mirror sources, $Q'(E)=0$, and for different
ratios of CMB temperatures $T'/T=0.1,\,0.3,\,0.5$. Observe a rather mild
dependence on generation spectral indexes between $\gamma_g = 2.3 \div
2.7$. }
\label{Qmir0}% 
\end{figure}

We take into account that intensity of cosmic ray sources may depend on
the cosmological redshift $z$ and parametrize the generation functions as 
\begin{equation} \label{Q} 
Q(E) \propto E^{-\gamma_g} (1+z)^m
\Theta(z_{\rm max} -z) \Theta(E_{\rm max} -E),
\end{equation} 
where $\gamma_g$ is a generation spectral index and $m$ is an evolution
parameter. We also assume that sources emerge at maximal redshift
$z_{\rm max}$ and their acceleration capacities are limited by energy $
E_{\rm max}$. For the mirror cosmic rays, we assume that their
generation function $Q'(E)$ has the same shape (\ref{Q}) as that of the
ordinary ones but the intensity can be different. In other words, we
take the ratio $Q'/Q$ to be constant. For the sake of definiteness, in
our following computations we take $m=3$, $z_{\rm max} = 4$ and $ E_{\rm
max}= 10$~ZeV. But we would like to stress that our results for the
UHECR spectral modification in the relevant energy range practically do
not depend on the choice of these parameters.

\begin{figure} % figure 2
\begin{minipage}{\columnwidth}
\centering
\includegraphics[width=\columnwidth]{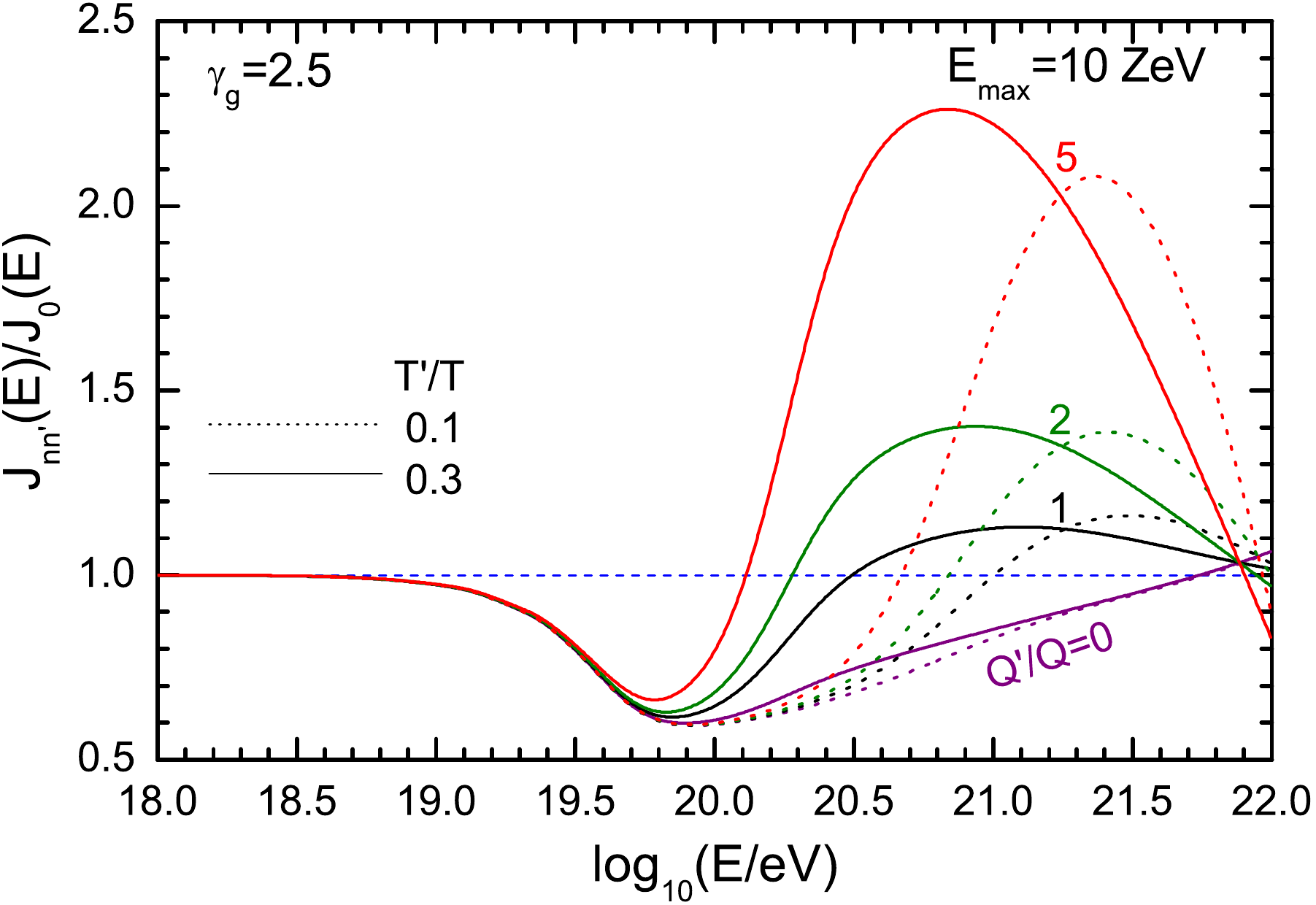} 
\end{minipage}
\caption{The same as on Fig.~\ref{Qmir0} but in the presence of mirror
sources with different intensities: $Q'(E)/Q(E)=0,\,1\,,2,\,5$. Observe
that the modification pattern in the cutoff region is practically
independent of the values of $T'/T$ and $Q'/Q$, but at the energies
above 100 EeV the dependence becomes very strong.} 
\label{Qmir}%
\end{figure}

\begin{figure} % figure 3
\begin{minipage}{\columnwidth}
\centering
\includegraphics[width=\columnwidth]{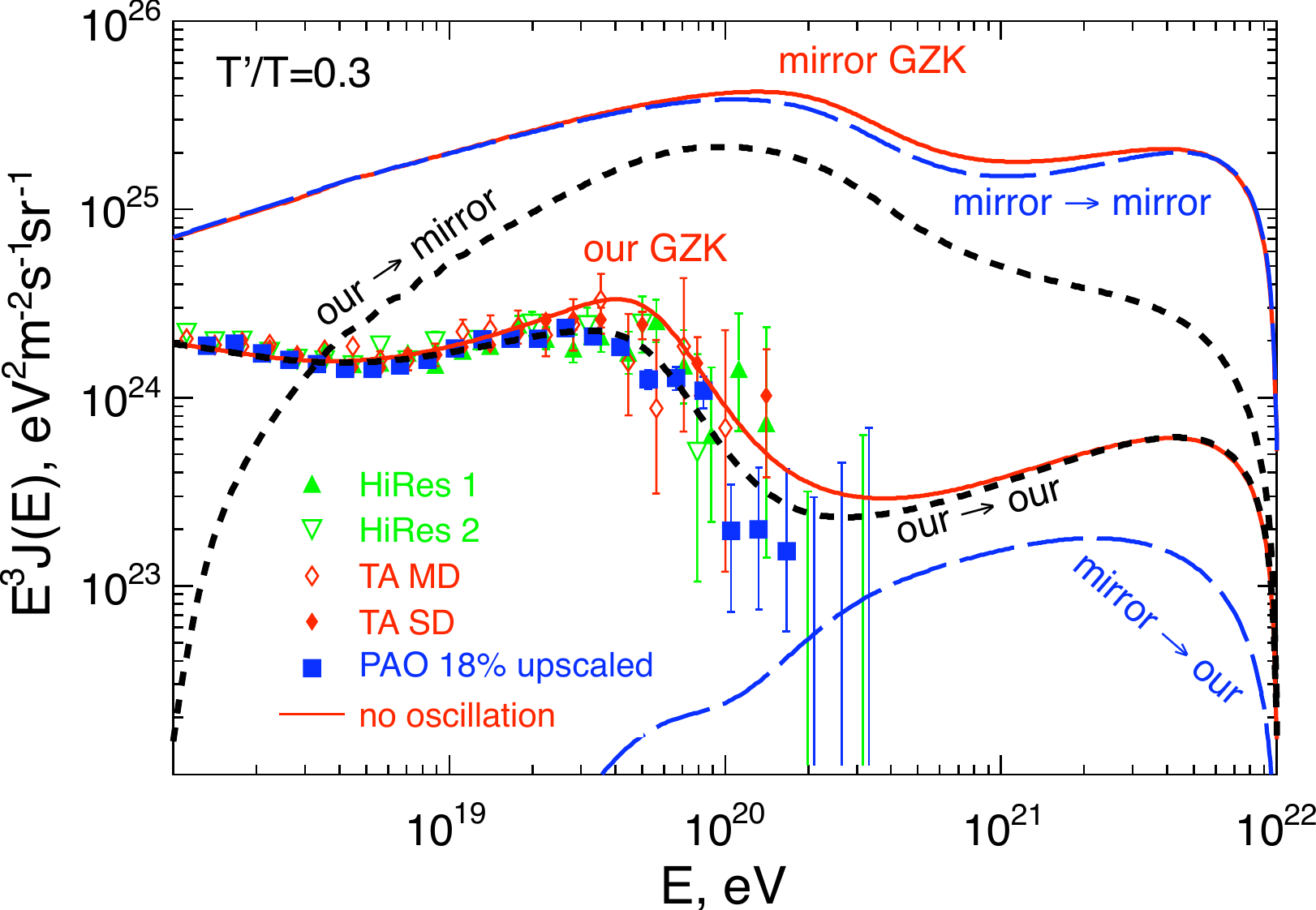}
\end{minipage}
\caption{The red solid curves marked as {\it our GZK} and {\it mirror
GZK} show the expected spectra of the ordinary and mirror cosmic rays in
the absence of $n$--$n'$ oscillation. The generation functions in two
sectors are taken identical, $Q(E)=Q'(E)$, with $\gamma_g=2.4$. We also
show experimental results of HiRes, TA and PAO (the latter are 18\%
upscaled in energy). The black short-dash curves show the spectra of
cosmic rays originated from ordinary sources $Q(E)$ that escape to
mirror sector due to $n$--$n'$ oscillation ({\it our$\to$mirror}) and
that remain in our sector ({\it our$\to$our}). Reciprocally, the blue
long-dash curves show mirror and ordinary cosmic ray fluxes originated
from mirror sources of the same intensity, $Q'(E)=Q(E)$. }
\label{All}% 
\end{figure}

The results of our calculations for different sets of parameters are
shown in Fig.~\ref{Qmir0} and Fig.~\ref{Qmir}. Let us first discuss the
case when there are no mirror cosmic rays sources, $Q'=0$. Then cosmic
rays with $E > E_{\rm GZK}$ produced in distant extragalactic sources
not only lose their energy during propagation, but also degrade in
number owing to $n$--$n'$ transition. This shifts the cutoff in the
cosmic ray spectrum to energies lower than $E_{\rm GZK} \simeq 60$~EeV.
Now the cutoff relates also to the non-conservation of the baryon
number: the most part of initial protons with $E > E_{\rm GZK}$
transforms into mirror protons, thus getting invisible for us, and so
the ordinary UHECR flux in the relevant energy range will be reduced
with respect to what is expected in the normal GZK scenario; on the
other hand, an imaginary mirror observer would detect a flux of mirror
UHECR originated in our world.\footnote{For demonstration, compare the
curves {\it our GZK}, {\it our$\to$our} and {\it our$\to$mirror} on
Fig.\ \ref{All} which result from real calculations for a concrete
choice of parameters ($\gamma_g=2.4$, $m=3$ and $T'/T=0.3$). Observe
that the ordinary sources induce a larger flux of the UHECR in mirror
sector than in our sector, even at energies $E < E_{\rm GZK}$. For the
UHECR produced at high redshift $z$, the $p\gamma \to N \pi$ reaction
threshold decreases by a factor $1+z$ while the respective m.f.p.
decreases by a factor $(1+z)^3$, which substantially increases energy
loses of the ordinary UHECR while for the UHECR that escape to mirror
sector the energy losses are much smaller.}
In other words, the baryon numbers B and B$'$ are not conserved
individually, but the sum B$+$B$'$ must be conserved.

In the Fig.~\ref{Qmir0} we show the modification factor $\eta(E)
=J_{nn'}(E)/J_0(E)$, defined as the ratio of UHECR spectra calculated
with and without $n-n'$ oscillation, in the absence of mirror sources,
$Q'=0$. Actually, it has an almost universal shape which just weakly
depends on the source generation function index $\gamma_g$ and on the
ratio of CMB temperatures $x=T'/T$ in two sectors. The spectral
modification due to oscillation starts earlier, at $E \sim 10$~EeV,
while the maximal difference from the GZK prediction (about 40$\%$) is
reached around $E=E_{\rm GZK}$.

Fig.~\ref{Qmir} shows the same spectral modification factor $\eta(E)$ in
the presence of the mirror sources, for different values of
$Q'(E)/Q(E)$. In fact, if mirror baryons constitute dark matter, one can
expect that there are also mirror cosmic rays. In fact, in quasars and
AGN the central black holes can be a potential sites for acceleration of
both ordinary and mirror protons. Therefore, it is natural to assume
that the ratio $Q'(E)/Q(E)$ is the same as the ratio of the relative
matter fractions, $\Omega'_B/\Omega_B$. Let us recall, that
$\Omega'_B/\Omega_B \sim 5$ can be naturally achieved in the joint
ordinary-mirror baryogenesis mechanism \cite{Bento:2001rc,Bento:2002sj}
provided that $x<0.3$ or so
\cite{Berezhiani:2003xm,Berezhiani:2005ek,Berezhiani:2008zza}.

In this case, in spite of larger m.f.p., the mirror cosmic rays with $E
> E'_{\rm GZK}$ can partially move to the ordinary sector which can
substantially increase the UHECR flux at energies above $E=100$~EeV
(compare the curves {\it mirror$\to$our} and {\it mirror$\to$mirror} on
Fig. \ref{All} which also shows that the GZK threshold energy in mirror
world is shifted as $E'_{\rm GZK}/E_{\rm GZK}\simeq T/T'$). The value of
the turning point depends on the parameter $x=T'/T$ as well as on the
strength of mirror sources, $Q'/Q$. Note however that the position of
the pre-GZK cutoff remains robust: the shape of the $\eta(E)$ at
$E=10-100$~EeV practically does not depend on the strength of the mirror
sources.\footnote{It is interesting to remark, that the fraction of
cosmic ray which originates in the parallel sector and then transforms
to ordinary cosmic rays due to $n$--$n'$ oscillation, can be composed by
anti-protons rather than by protons, depending on the sign of the baryon
asymmetry of the parallel world. However, in the high energy cosmic ray
showers it is difficult to distinguish experimentally primary protons
from antiprotons. }

The earlier end of the cosmic ray spectrum seems to be indicated by the
data of Pierre Auger Observatory (PAO). The PAO data show the cutoff
starting from $E \simeq 25$~EeV \cite{2011arXiv1107.4809T}, a factor of
2 lower than $E_{\rm GZK}$. The estimated systematic uncertainty of $22
\%$ in the energy definition is not sufficient for adjusting the cutoff
position to the GZK shape. In Fig. \ref{All} we show the PAO spectrum
are upscaled by 18\% with respect to original data reported in
\cite{2011arXiv1107.4809T}. (This is compatible with the 22\% margins
allowed by PAO for systematic errors.) Such upscaling in energies, from
the one side, renders the PAO data compatible with the data accumulated
by other experiments HiRes \cite{Sokolsky:2011zz} and TA
\cite{AbuZayyad:2012ru}, in the energy range $E < 20$ EeV where all
experiments reached very good accuracy, and on the other side, with the
spectral characteristics predicted by dip model that considers protons
as the cosmic ray carriers \cite{Berezinsky:2002nc}. One can immediately
observe that the standard UHECR spectrum (the red solid curve {\it our
GZK} on Fig.\ \ref{All}), while is perfectly compatible with the PAO
data at $E < 20$ EeV but gets into evident conflict at the energies
above $20$ EeV.\footnote{ Needless to say, for the original PAO data
\cite{2011arXiv1107.4809T}, without being upscaled by 18\%, disagreement
with the GZK spectrum would become stronger. }
On the other hand, our result which takes into account $n$--$n'$
oscillation (the black short-dash curve {\it our$\to$our} on Fig.
\ref{All}) matches the PAO data much better; in particular, it well
reflects the stiffening of the spectrum from $E \simeq 30$ EeV or so
\cite{2011arXiv1107.4809T}. As for the HiRes and TA, their data at $E >
30$ EeV are rather scattered and have large error bars which in fact
renders them compatible with the standard GZK prediction as well as to
its $n$--$n'$ modification. In fact, the $n$-$n'$ oscillation model
proposed in this Letter mitigates the controversy between the data of
PAO and the data of HiRes and TA experiments putting the beginning of
the cutoff in the middle, when extragalactic protons are assumed to be
the cosmic ray carriers.

However, we consider that it would be premature to claim that the
problem is solved until the experimental situation is not well settled.
A controversy concerning the shape and chemical composition of the UHECR
spectrum between the PAO data from one side, and the data of HiRes
\cite{Sokolsky:2011zz} and TA \cite{AbuZayyad:2012ru} on the other side,
is not yet resolved (see e.g.\ \cite{Aloisio:2009sj}). The HiRes and TA
data indicate cosmic rays to be protons with cutoff at $E_{\rm GZK}$,
but their statistic is lower. On the other hand, the PAO data are
consistent to protons up to $E\simeq 5-10$~EeV, but disfavor the protons
at larger energies at which the UHECR spectrum seems to be dominated by
nuclei. Moreover, it seems that the nuclei must become gradually heavier
with increasing the energy which seems very controversial in itself. It
is worth to note, however, that the issue of the composition determining
at such high energies is a subject of theoretical models rather than of
experimentally measured cross-section and multiplicities at the relevant
energies. It cannot be excluded that the distribution pattern of the
shower maxima observed by the PAO is not due to appearance of the nuclei
starting from the energies of about 5~EeV but rather points to new
`strong' physics at the energy scales $\sqrt{s} > 10$~TeV, perhaps the
same one which is also at the origin of the baryon number violating
interactions as e.g.\ $(1/M)^5(udd)(u'd'd')$ that generates $n$--$n'$
mixing itself. It should be noted, that in case the UHECR are mostly
heavy nuclei, the neutrons produced in their photodisintegration on CMB
will be also lost as they escape to the parallel world due to $n$--$n'$
oscillation. We think that more data with higher accuracy must be
collected by the currently operating PAO and TA experiments for settling
the situation which would allow to critically test also our hypothesis.

Prompt neutron oscillations may provide correlation of cosmic rays with
$E > 100$ EeV with distant sources (e.g.\ BL Lacs
\cite{Tinyakov:2001nr,Tinyakov:2001ir}) as an indication of the UHECR
transport in the parallel world. It is natural to assume that BL Lacs
and blasars in general can be the natural sites for acceleration of both
ordinary and mirror cosmic rays, as far as the central black holes
considered to be the acceleration engines must be democratic for both
types of the matter. As far as the CMB of the latter is much colder than
ours, the mirror UHECR from the distant sources at several hundred Mpc
from us could arrive practically without losing their energy and can be
converted to ordinary cosmic rays via $p'\gamma' \to n' \pi'$ scattering
within our GZK radius and subsequent conversion $n'\to n$ and decay
$n\to p e \nu$.

Another immediate consequence of the baryon losses due $n-n'$
transitions is a strong suppression of the cosmogenic
Berezinsky-Zatsepin (BZ) \cite{Berezinsky:1969zz} neutrino flux mostly
produced via $p\gamma \rightarrow n\pi^+$ scattering with following
$\pi^+ \rightarrow \mu^+ \nu_\mu$ and $\mu^+ \rightarrow e^+ 
\nu_e + \bar{\nu}_\mu$ decays. This conclusion remains pessimistic 
even if ordinary and mirror neutrinos also have non-zero mixing
\cite{Foot:1995pa,Berezhiani:1995yi}. Really, since $T' < T$, mirror
cosmic rays have very large m.f.p., $l'_{s} \sim x^{-3} l_{s} $ (e.g.\
$l'_{s} \sim 600$~Mpc for $x=0.2$ versus $l_{s}\sim 5$ Mpc), and thus
suffer much less scatterings than ordinary ones. Therefore the diffuse
cosmogenic neutrino flux may turn out to be much lower than expected
\cite{Berezinsky:2010xa}. Even the giant ICECUBE \cite{Abbasi:2011ji}
with its control over $1$ km$^3$ of ice may be insufficient to detect
this flux.

The electromagnetic cascades originated from $p\gamma \rightarrow
p\pi^0$ channel with subsequent $\pi^0 \rightarrow 2\gamma$ decay and
$\gamma \gamma_{\rm CMB} \rightarrow e^+ e^-$ will be also suppressed,
since most amount of the ordinary super-GZK cosmic rays escape to the
parallel world just after few proton scatterings off CMB. In this way,
one can soften also the restrictions \cite{Berezinsky:2010xa} imposed on
the UHECR models by the diffuse extragalactic gamma-ray flux measured by
Fermi-LAT at $E \gtrsim 100$ GeV \cite{Abdo:2010nz}.

To conclude, concept of parallel/mirror sector with exactly the same
microphysics as the ordinary particle sector provides an interesting
possibility for dark matter
\cite{Berezhiani:2000gw,Ignatiev:2003js,Berezhiani:2003wj,Berezhiani:2005vv,Bento:2001rc,Bento:2002sj,Berezhiani:2003xm,Berezhiani:2005ek,Berezhiani:2008zza} 
and a peculiar way for its testing via oscillation
phenomena of ordinary neutral particles in their mass degenerate mirror
twins. In particular, the effects of fast neutron -- mirror neutron
oscillation may be detected in the laboratory conditions. The present
situation is very intriguing in view of the experimental signal for the
anomalous neutron loses which can be explained by $n$--$n'$ oscillation
with a timescale of few seconds \cite{Berezhiani:2012rq}. If future high
accuracy experiments on the neutron disappearance and regeneration will
eventually confirm the claim, then deepest consequences will follow for
particle physics, astrophysics and cosmology. Namely, the underlying
TeV-scale physics can be tested at the LHC, and it can have profound
cosmological implications for the primordial baryogenesis and dark
matter \cite{Berezhiani:2005hv,Berezhiani:2006je,Berezhiani:2008bc}.
Here we show that this phenomenon may complementary lead to interesting
astrophysical consequences for the cosmic ray spectrum at the GZK
region. Due to the baryon non-conservation during the UHECR propagation,
the cutoff of the spectrum shifts to lower energies and becomes
significantly steeper. Such a spectral modification will be testable at
the operating detectors with further increase of the statistics and
clearing the controversy regarding the chemical composition. On the
other hand, in the presence of powerful mirror sources the UHECR
spectrum at highest energies may get even higher than one expects in the
standard GZK case. Unfortunately, the data statistics accumulated at the
operating installations is not yet enough to determine the exact shape
of the spectrum above $100$~EeV. But this can be achieved at the future
JEM-EUSO mission \cite{Takahashi:2009zzc}.

\vspace{2mm} 

\noindent
{\bf Acknowledgements.}
The work of Z.B. is partially supported by the MIUR grant PRIN'08 on
Astroparticle Physics. A.G. thanks the LNGS for hospitality during the
initial period of this work.

\end{document}